\documentclass[letterpaper,12pt]{article}
\usepackage[utf8]{inputenc}
\usepackage[T1]{fontenc}
\usepackage{lmodern}
\usepackage[margin=1.0in]{geometry}
\usepackage[numbers,super]{natbib}
\usepackage{mathtools,amsmath,amssymb}
\usepackage{microtype}
\usepackage{appendix}
\usepackage[ruled]{algorithm2e}
\usepackage{hyperref}
\usepackage{setspace}      
\allowdisplaybreaks

\begin{document}

\title{Variational phase recovering without phase unwrapping in phase-shifting interferometry}
\author{Ricardo Legarda-Saenz$^{\text{a}}$, Alejandro Téllez Quiñones$^{\text{b}},$ Carlos Brito-Loeza$^{\text{a}},$\\ 
Arturo Espinosa-Romero$^{\text{a}}$\\
$^{\text{a}}${\small CLIR at Facultad de Matemáticas, Universidad Autónoma de Yucatán.}\\
{\small Anillo Periférico Norte, Tablaje Cat. 13615.}\\
{\small C.P. 97203, Mérida, Yucatán. México}\\
$^{\text{b}}${\small CONACYT-Centro de Investigación en Geografía y Geomática, Unidad Mérida.}\\
{\small Carretera Sierra Papacal-Chuburna Puerto Km 5.}\\
{\small C.P. 97302, Sierra Papacal-Yucatán. México}\\
E-mail: \texttt{rlegarda@correo.uady.mx}    }
\date{\today}

\maketitle

\begin{abstract}
We present a variational method for recovering the phase term from the information obtained from phase-shifting methods. First we introduce the new method based on a variational approach and then describe the numerical solution of the proposed cost function, which results in a simple algorithm. Numerical experiments with both synthetic and real fringe patterns shows the accuracy and simplicity of the resulting algorithm.
\end{abstract}


\section{Introduction}
The main goal of fringe analysis techniques is to recover accurately the modulated phase from one or several fringe patterns~\cite{Robinson1993,Rajshekhar2012}; such phase is related to some physical quantities like shape, deformation, refractive index, temperature, etc. The basic model for a fringe pattern is given by \[I_{\mathbf{x}} = a_{\mathbf{x}} + b_{\mathbf{x}}\cos\left(\phi_{\mathbf{x}}\right),\] where $\mathbf{x} = (x,y)$, $a_{\mathbf{x}}$ is the background illumination, $b_{\mathbf{x}}$ is the amplitude modulation, and $\phi_{\mathbf{x}}$ is the phase map to be recovered.

Among the methods for phase estimation is the phase-shifting method~\cite{Gasvik2002,Servin2014}, which consists in acquiring several fringe patterns where the phase term is incremented between successive frames. Such fringe patterns are  defined as \[I_{\mathbf{x},k} = a_{\mathbf{x}} + b_{\mathbf{x}}\cos\left(\phi_{\mathbf{x}} + \alpha_{k}\right),\; k = 1, \ldots,K, \; K\geq 3,\] where $\alpha_{k}$ is the phase step and $K$ is the number of fringe patterns used. For every $\alpha_{k}$, the fringe pattern can be written as
\begin{equation}\label{eq:phaseShift}
I_{\mathbf{x},k} = a_{\mathbf{x}} + b_{\mathbf{x}}\cos\left(\phi_{\mathbf{x}} + \alpha_{k}\right)
= I^{0}_{\mathbf{x}} + I^{c}_{\mathbf{x}}\cos\left(\alpha_{k}\right) - I^{s}_{\mathbf{x}}\sin\left(\alpha_{k}\right)
\end{equation}
where
\begin{align*}
I^{0}_{\mathbf{x}} &= a_{\mathbf{x}},\\
I^{c}_{\mathbf{x}} &= b_{\mathbf{x}}\cos\left(\phi_{\mathbf{x}}\right),\\
I^{s}_{\mathbf{x}} &= b_{\mathbf{x}}\sin\left(\phi_{\mathbf{x}}\right).
\end{align*}
The funtions $I^{0}_{\mathbf{x}},I^{c}_{\mathbf{x}},I^{s}_{\mathbf{x}}$ can be estimated with different phase-shifting techniques~\cite{Tellez-Quinones2010,Tellez-Quinones2012,Yatabe2016b,Yatabe2017a,Yatabe2017}. Using these coefficients, the wrapped phase term can be computed by~\cite{Malacara2005} 
\begin{equation}\label{eq:atan}
\hat{\phi}_{\mathbf{x}} = \mathtt{atan2}\left[I^{s}_{\mathbf{x}},I^{c}_{\mathbf{x}}\right] = \phi_{\mathbf{x}}\,\bmod{2\pi}.
\end{equation}
Then the phase term $\phi_{\mathbf{x}}$ is estimated by a process named phase unwrapping~\cite{Ghiglia1998}, which is usually computationally intensive and susceptible to noise. To avoid these drawbacks, several approaches have been developed to estimate the phase without the need of the unwrapping process~\cite{Servin1997a,Legarda-Saenz2002a,Legarda-Saenz2014,Yatabe2016}.

A different approach is found in reference \citenum{Paez1999}, where the information obtained from the phase-shifting method, given in Eq. (\ref{eq:phaseShift}), is used to estimate the phase term $\phi_{\mathbf{x}}$ from the information of $I^{c}_{\mathbf{x}}$ and $I^{s}_{\mathbf{x}},$ and the partial derivatives of these two functions, respectively. This approach consists in computing the gradient field as follows:
\begin{equation}\label{eq:gradiente}
\begin{aligned}
\Phi_{\mathbf{x}} &= \left( \Phi_{\mathbf{x}}^{1},\; \Phi_{\mathbf{x}}^{2} \right) = \left( \frac{\partial\phi_{\mathbf{x}}}{\partial x},\; \frac{\partial\phi_{\mathbf{x}}}{\partial y} \right)\\
&= \left( \frac{\frac{\partial I^{s}_{\mathbf{x}}}{\partial x}\,I^{c}_{\mathbf{x}}\;-\;I^{s}_{\mathbf{x}}\,\frac{\partial I^{c}_{\mathbf{x}}}{\partial x}}{\left(I^{c}_{\mathbf{x}}\right)^{2} + \left(I^{s}_{\mathbf{x}}\right)^{2}},  \;  \frac{\frac{\partial I^{s}_{\mathbf{x}}}{\partial y}\,I^{c}_{\mathbf{x}}\;-\;I^{s}_{\mathbf{x}}\,\frac{\partial I^{c}_{\mathbf{x}}}{\partial y}}{\left(I^{c}_{\mathbf{x}}\right)^{2} + \left(I^{s}_{\mathbf{x}}\right)^{2}} \right).
\end{aligned}
\end{equation}
From the gradient field $\Phi_{\mathbf{x}}$, the phase term $\phi_{\mathbf{x}}$ is estimated by using line integrals~\cite{Paez1999}. In this way, the non-linearity of the arctangent function, Eq. (\ref{eq:atan}), is avoided. This approach has been successfully used on the demodulation of fringe patterns obtained from phase-shifting methods~\cite{Tellez-Quinones2012,Tellez-Quinones2012b}. However, the line integrals approach fails with moderate levels of noise and/or aliasing in the input fringe patterns, in the same way the Itoh's method does~\cite{Ghiglia1998}.

In this work we present a method for recovering the phase term $\phi_{\mathbf{x}}$ from the information obtained from phase-shifting methods; that is, using only the fringe patterns $I^{c}_{\mathbf{x}}$ and $I^{s}_{\mathbf{x}}$, avoiding the use of the nonlinear arctangent function. First we introduce the new method based on a variational approach. Then we describe the numerical solution of the proposed cost function, which results in a simple algorithm. The performance of the proposed method is evaluated by numerical experiments with both synthetic and real data. A comparison against two well-known least-square based unwrapping methods is also presented. Finally we discuss our results and present some concluding remarks.

\section{A new variational model for the recovery of the phase from phase-shifting method}

\subsection{Variational Formulation}
Variational techniques have been successfully used in fringe pattern processing. In the literature it is possible to find several works about fringe-pattern filtering~\cite{Zhang2008,Zhu2013b,Li2017b}, demodulation~\cite{Legarda-Saenz2014}, unwrapping~\cite{Lacombe2002} and gradient-field estimation of a wrapped-phase for unwrapping processing~\cite{Huang2012}. 

In this work, we propose to estimate the phase map $\phi_{\mathbf{x}}$ as the solution of the minimization problem defined by
\begin{equation}\label{eq:energiaMinimiza}
\underset{\phi}{\min}\; E(\phi_{\mathbf{x}}),
\end{equation}
where
\begin{align*}
E(\phi_{\mathbf{x}}) &= \frac{1}{2}\int_{\Omega}\left|\nabla\phi_{\mathbf{x}} - \Phi_{\mathbf{x}}\right|^{2}d\mathbf{x} 
+ \frac{1}{2}\int_{\Omega}\left(b_{\mathbf{x}}\cos\phi_{\mathbf{x}} - I^{c}_{\mathbf{x}}\right)^{2}d\mathbf{x}\\ 
&+ \frac{1}{2}\int_{\Omega}\left(b_{\mathbf{x}}\sin\phi_{\mathbf{x}} - I^{s}_{\mathbf{x}}\right)^{2}d\mathbf{x} 
+\frac{\lambda}{2}\int_{\Omega}\left|\nabla\phi_{\mathbf{x}}\right|^{2}d\mathbf{x}
\end{align*}
and $I^{s}_{\mathbf{x}},$ $I^{c}_{\mathbf{x}}$ are the input fringe patterns obtained from the phase-shifting method, given in Eq. (\ref{eq:phaseShift}); the term $b_{\mathbf{x}}$ is estimated from these fringe patterns in the following way \[b_{\mathbf{x}} = \sqrt{\left( I^{s}_{\mathbf{x}}\right)^{2} + \left( I^{c}_{\mathbf{x}}\right)^{2}}.\] The term $\Phi_{\mathbf{x}}$ is the gradient field estimated from the input fringe patterns described in Eq. (\ref{eq:gradiente}), $\Omega \subset \mathbb{R}^2$ denotes the continuous signal domain, and $\lambda > 0$ is a Lagrange multiplier. 

The motivation to our proposed model is two fold. First, we note that the first term in Eq.(\ref{eq:energiaMinimiza}) penalizes the differences between the known and possibly noisy gradient phase field and the recovered gradient phase field, and this term can be seen as an equivalent expression of the least-square approach to the phase unwrapping technique described in reference \citenum{Ghiglia1998}. Also, by the action of the last term, the recovered phase field will be smoothed and the problem made well-posed. By just using these two terms, there will be many possible solutions since the gradient field of more than one phase surface will be a feasible solution. In order to avoid this problem, we inserted the second and third terms in Eq.(\ref{eq:energiaMinimiza}), to enforce the solution to be close to the input information, see Eqs. (\ref{eq:phaseShift}) and (\ref{eq:gradiente}). In that sense, our model is more robust than similar models used in unwrapping processes that lack a way to constrain the solution to the expected scale and shape.

To obtain the solution of the problem expressed in Eq.(\ref{eq:energiaMinimiza}), the first order optimality condition or Euler-Lagrange equation has to be derived. In the formal derivation we assume that the function $\phi_{\mathbf{x}}$ is smooth enough such that gradients are well defined and the variation $\delta \phi_{\mathbf{x}}$ has compact support over $\Omega$ so that we can use the divergence theorem to get rid of the boundary term. 

To simplify notation, write $\left\langle f \right\rangle = \int_{\Omega} f\,d\mathbf{x}$ or $\left\langle f \right\rangle_{\partial} = \int_{\partial\Omega}f\,d\mathbf{x}$ depending on whether the integral is evaluated on the domain $\Omega$ or its boundary $\partial\Omega$. Then the first variation is derived as
\begin{equation}\label{eq:energiaVariacional}
\begin{aligned}
\delta E(\phi_{\mathbf{x}}) &= \frac{1}{2}\left\langle \delta\big|\nabla\phi_{\mathbf{x}} - \Phi_{\mathbf{x}}\big|^{2} \right\rangle 
+ \frac{1}{2}\left\langle \delta\big(b_{\mathbf{x}}\cos\phi_{\mathbf{x}} - I^{c}_{\mathbf{x}}\big)^{2} \right\rangle\\ 
&+ \frac{1}{2}\left\langle \delta\big(b_{\mathbf{x}}\sin\phi_{\mathbf{x}} - I^{s}_{\mathbf{x}}\big)^{2} \right\rangle 
+ \frac{\lambda}{2}\left\langle \delta\big|\nabla\phi_{\mathbf{x}}\big|^{2} \right\rangle \\ 
&= \frac{1}{2}\left\langle 2\big(\nabla\phi_{\mathbf{x}} - \Phi_{\mathbf{x}}\big)\cdot\delta\nabla\phi_{\mathbf{x}} \right\rangle 
+ \frac{1}{2}\left\langle 2\big(b_{\mathbf{x}}\cos\phi_{\mathbf{x}} - I^{c}_{\mathbf{x}}\big)\cdot\delta\left(b_{\mathbf{x}}\cos\phi_{\mathbf{x}}\right)  \right\rangle\\ 
&+ \frac{1}{2}\left\langle 2\big(b_{\mathbf{x}}\sin\phi_{\mathbf{x}} - I^{s}_{\mathbf{x}}\big)\cdot\delta\left(b_{\mathbf{x}}\sin\phi_{\mathbf{x}}\right) \right\rangle 
+ \frac{\lambda}{2}\left\langle 2\nabla\phi_{\mathbf{x}}\cdot\delta\nabla\phi_{\mathbf{x}} \right\rangle\\ 
&= \left\langle \delta\phi_{\mathbf{x}}\big(\nabla\phi_{\mathbf{x}}-\Phi_{\mathbf{x}}\big)\cdot\mathbf{n} \right\rangle_{\partial} - \left\langle \nabla\cdotp\big(\nabla\phi_{\mathbf{x}} - \Phi_{\mathbf{x}}\big)\;\delta\phi_{\mathbf{x}} \right\rangle\\ 
&+ \left\langle \big(b_{\mathbf{x}}\cos\phi_{\mathbf{x}} - I^{c}_{\mathbf{x}}\big)\cdotp\big(-b_{\mathbf{x}}\sin\phi_{\mathbf{x}}\big)\;\delta\phi_{\mathbf{x}} \right\rangle\\ 
&+ \left\langle \big(b_{\mathbf{x}}\sin\phi_{\mathbf{x}} - I^{s}_{\mathbf{x}}\big)\cdotp\big(b_{\mathbf{x}}\cos\phi_{\mathbf{x}}\big)\;\delta\phi_{\mathbf{x}}\right\rangle\\
&+ \lambda\left\langle \delta\phi_{\mathbf{x}}\nabla\phi_{\mathbf{x}}\cdot\mathbf{n} \right\rangle_{\partial} - \lambda\left\langle \big(\nabla\cdotp\nabla\phi_{\mathbf{x}}\big)\;\delta\phi_{\mathbf{x}} \right\rangle\\
\end{aligned}
\end{equation}
where in the last line we have used the divergence theorem and $\mathbf{n}$ denotes the unit outer normal vector to the boundary. Finally, the variational derivative of $E(\phi_{\mathbf{x}})$ is given by
\begin{equation}\label{eq:gradienteE}
\begin{aligned}
\frac{\partial E(\phi_{\mathbf{x}})}{\partial\phi_{\mathbf{x}}} &= -\nabla\cdotp\big(\nabla\phi_{\mathbf{x}} - \Phi_{\mathbf{x}}\big) + \big(b_{\mathbf{x}}\cos\phi_{\mathbf{x}} - I^{c}_{\mathbf{x}}\big)\cdotp\big(-b_{\mathbf{x}}\sin\phi_{\mathbf{x}}\big)\\ 
&+ \big(b_{\mathbf{x}}\sin\phi_{\mathbf{x}} - I^{s}_{\mathbf{x}}\big)\cdotp\big(b_{\mathbf{x}}\cos\phi_{\mathbf{x}}\big) -\lambda\nabla\cdotp\nabla\phi_{\mathbf{x}}\\
&= -\left(1 + \lambda\right)\nabla\cdotp\nabla\phi_{\mathbf{x}} + \nabla\cdotp\Phi_{\mathbf{x}}
+I^{c}_{\mathbf{x}}\cdotp\big(b_{\mathbf{x}}\sin\phi_{\mathbf{x}}\big) - I^{s}_{\mathbf{x}}\cdotp\big(b_{\mathbf{x}}\cos\phi_{\mathbf{x}}\big) = 0
\end{aligned}
\end{equation}
with boundary conditions
\begin{equation}
\begin{aligned}
\big(\nabla\phi_{\mathbf{x}} - \Phi_{\mathbf{x}}\big)\cdot\mathbf{n} &= 0\\
\nabla\phi_{\mathbf{x}}\cdot\mathbf{n} &= 0
\end{aligned}
\end{equation}

\subsection{Numerical Solution}
Let $\phi_{i,j}=\phi(x_{i},y_{j})$ to denote the value of a grid function $\phi$ at point $(x_{i},y_{j})$ defined on $\Omega = [a,b]\times[c,d]$ where the sampling points of the grid are 
\begin{align*}
x_{i} &= a + (i-1)h_x\\
y_{j} &= c + (j-1)h_y
\end{align*}
with $1\leq i\leq m,\;1\leq j\leq n,$ and $h_x=(b-a)/(m-1), h_y=(d-c)/(n-1)$. 

To approximate the derivatives, we use central finite differences between ghost half-points as follows
\[\delta_{x}\phi_{i,j} = \frac{\phi_{i+1/2,j}-\phi_{i-1/2,j}}{h_x},\quad\text{and}\quad
\delta_{y}\phi_{i,j} = \frac{\phi_{i,j+1/2}-\phi_{i,j-1/2}}{h_y}.\] 
The divergence term in Eq. (\ref{eq:gradienteE}) is approximated as \[\nabla \cdot V_{i,j} = \delta_{x}V^1_{i,j} + \delta_{y}V^2_{i,j},\] where
\begin{align*}
V_{i,j} &= (V^1_{i,j},V^2_{i,j}) = \nabla\phi_{i,j} - \Phi_{i,j}, \\
\nabla \phi_{i,j} &= (\delta_{x}\phi_{i,j}, \delta_{y}\phi_{i,j}).
\end{align*}
The rest of the terms in the equation are approximated by straight forward evaluation at point $(x_{i},y_{j})$. 

To implement the boundary condition on $\partial \Omega$, we assume without loss of generality that $\mathbf{n} = (\pm 1, 0)$ and $\mathbf{n} = (0,\pm 1)$ in the $x-$ and $y-$direction, respectively. With this consideration, the first boundary condition is expressed as  
\begin{equation}
\begin{aligned}
\delta_{x}\phi_{m,j} - \Phi_{m,j}^{1} &= 0\quad\text{for}\quad \mathbf{n} = (1,0),\\
-\left(\delta_{x}\phi_{1,j} - \Phi_{1,j}^{1} \right)  &= 0\quad\text{for}\quad \mathbf{n} = (-1,0),\\
\delta_{y}\phi_{i,n} - \Phi_{i,n}^{2} &= 0\quad\text{for}\quad \mathbf{n} = (0,1),\\
-\left(\delta_{y}\phi_{i,1} - \Phi_{i,1}^{2} \right)  &= 0\quad\text{for}\quad \mathbf{n} = (0,-1).\\
\end{aligned}
\end{equation}
Examples of numerical implementations of similar functionals can be found on references \citenum{Rudin1992} and \citenum{Getreuer2012}.

\section{Numerical Experiments}
To illustrate the performance of the proposed method, we carried out some numerical experiments using a Intel Core i7 @ 2.40 GHz laptop with Debian GNU/Linux 8 (jessie) 64--bit and 16 GB of memory. For both experiments, we solve Eq. (\ref{eq:gradienteE}) using a fast variant of Nesterov’s method, which is an improvement of the gradient descent method~\cite{Kim2016a,ODonoghue2015}. In our experiments we found that the Nesterov’s method is approximately 58 times faster than the gradient descent method, as far as iterations is concerned. This method is given by
\begin{equation}\label{eq:Nesterov}
\begin{aligned}
\beta^{k+1}_{\mathbf{x}} &= \phi^{k}_{\mathbf{x}} - \tau\;\frac{\partial E(\phi_{\mathbf{x}})}{\partial\phi_{\mathbf{x}}},\\ 
t^{k+1} &=  \frac{1+ \sqrt{1 + 4\left(t^{k}\right)^{2}}}{2},\\ 
\phi^{k+1}_{\mathbf{x}} &= \beta^{k+1}_{\mathbf{x}} + \frac{t^{k}-1}{t^{k+1}}\left(\beta^{k+1}_{\mathbf{x}} - \beta^{k}_{\mathbf{x}}\right) + \frac{t^{k}}{t^{k+1}}\left(\beta^{k+1}_{\mathbf{x}} - \phi^{k}_{\mathbf{x}}\right),
\end{aligned}
 \end{equation} 
where $\beta^{0}_{\mathbf{x}} = \phi^{0}_{\mathbf{x}},$ $t^{0} =  1$, $k = 0, 1,2,\ldots,$ and $\tau$ is the step size of the gradient descent. We chose the step size $\tau$ using the algorithm proposed in reference \citenum{Shi2005}, which estimates the Lipschitz constant of the functional.

We use as stopping criteria for our optimization algorithm the same terms used in reference \citenum{Vogel1996} with $\delta_1=\delta_2 = \delta_3 = 10^{-7}$ and $k_{max} = 15000.$ For simplicity, we selected the regularization parameter $\lambda$ manually; however, well known methods can be used to obtain the best parameter for this task, such as those described in section 5.6 of reference \citenum{Bertero1998}. In addition, we use a normalized error $Q$ to compare the phase-map estimation; this error is defined as~\cite{Perlin2016}:
\begin{equation}
Q\left(\mu,\nu\right) = \frac{\| \mu - \nu\|_{2}}{\| \mu\|_{2} + \|\nu\|_{2}}, 
\end{equation}
where $\mu$ and $\nu$ are the signals to be compared. The normalized error values vary between zero (for perfect agreement) and one (for perfect disagreement).

\subsection{Phase estimation using synthetic fringe patterns}
\begin{figure}[ht!]
  \begin{center}
    \includegraphics[width=\textwidth]{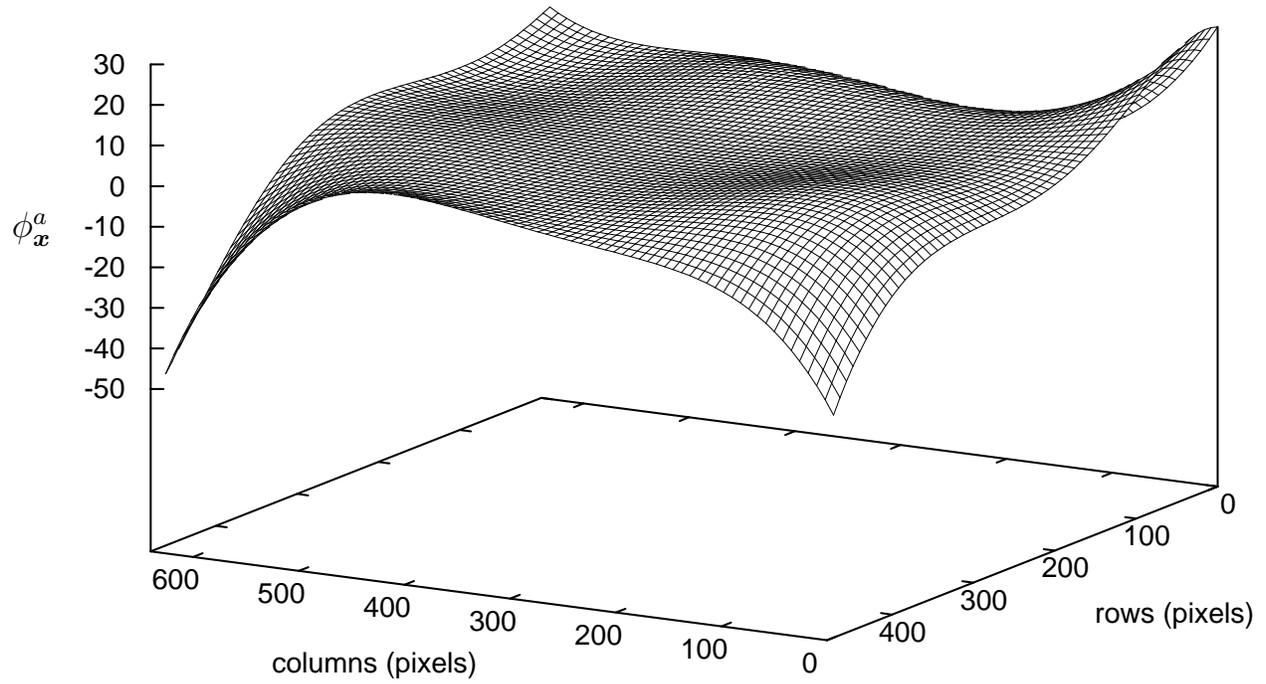}
    \caption{Synthetic phase map given in Eq. (\ref{eq:wavefront}).}
    \label{fig:S1phaseMap}
  \end{center}
\end{figure}

\begin{figure}[ht!]
  \begin{center}
    \includegraphics[width=\textwidth]{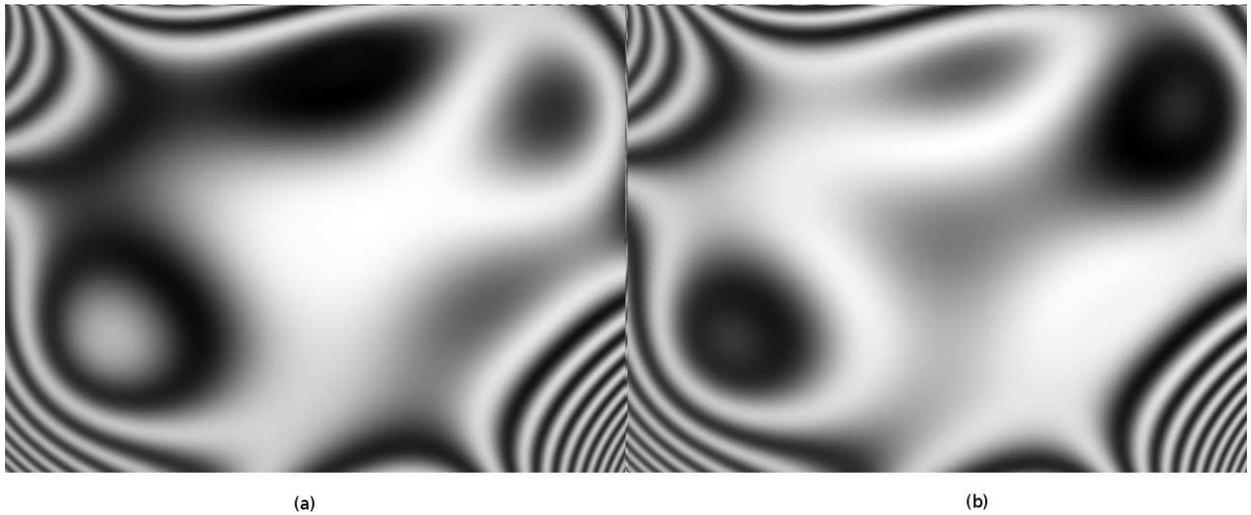}
    \caption{Fringe patterns generated with the phase map given in Eq. (\ref{eq:wavefront}): (a) $I^{c}_{\mathbf{x}}$, (b) $I^{s}_{\mathbf{x}}$.}
    \label{fig:S1fringes}
  \end{center}
\end{figure}

\begin{figure}[ht!]
  \begin{center}
    \includegraphics[width=\textwidth]{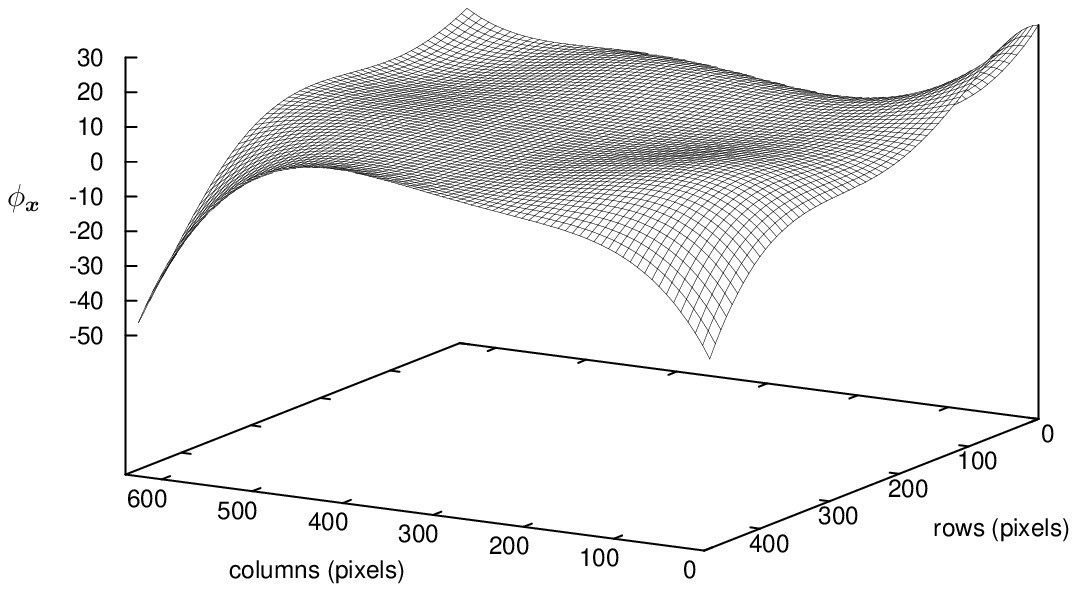}
    \caption{Estimated phase map using fringe patterns shown in Fig. \ref{fig:S1fringes}.}
    \label{fig:S1estimacion}
  \end{center}
\end{figure}

\begin{figure}[ht!]
  \begin{center}
    \includegraphics[width=\textwidth]{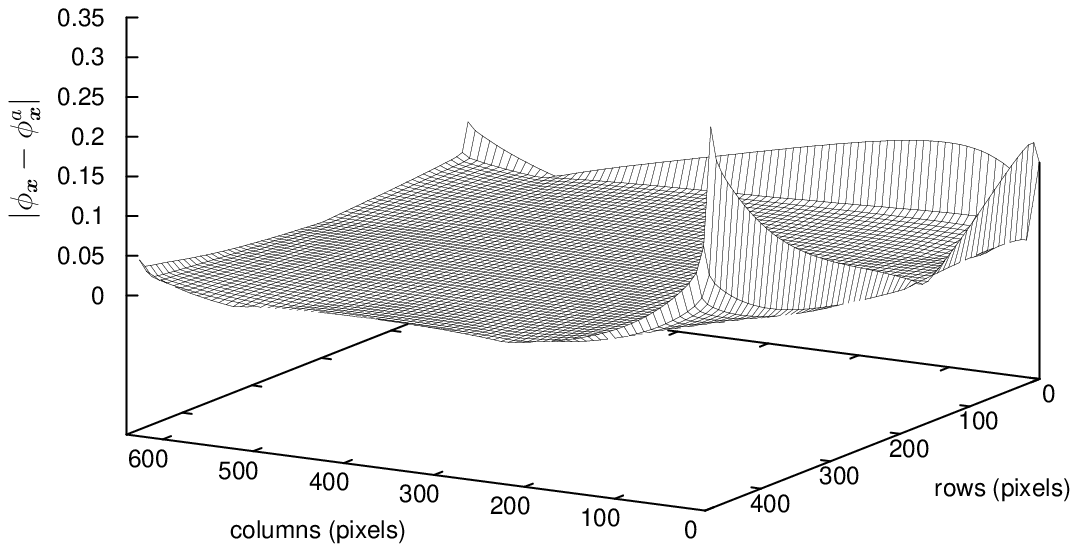}
    \caption{Absolute difference between the estimated phase map using fringe patterns shown in Fig. \ref{fig:S1fringes} and the synthetic phase map given in Eq. (\ref{eq:wavefront}).}
    \label{fig:S1errorEst}
  \end{center}
\end{figure}

The first set of experiments was the estimation of a synthetic phase map defined as~\cite{Tellez-Quinones2012}:
\begin{equation}
\begin{aligned}\label{eq:wavefront}
\phi_{\mathbf{x}}^{a} &= 1.3 - 1.9x 
- 1.3\left(1 - 6y^{2} - 6x^{2} + 6y^{4} + 12x^{2}y^{2} + 6x^{4}\right)\\ 
&+ 3.415\left(5xy^{4} - 10x^{3}y^{2} + x^{5}\right)\\
&+ 0.43\left(3x - 12xy^{2} - 12x^{3} + 10xy^{4} + 20x^{3}y^{2} + 10x^{5}\right)\\
&+ 2.6\left(-4y^{3} + 12x^{2}y + 5y^{5} - 10x^{2}y^{3} - 15x^{4}\right), \\
\end{aligned}
\end{equation}
evaluated in a square domain $\Omega = \left\lbrace (x,y)\,|\, -1\leq x,y\leq 1\right\rbrace.$ Figure \ref{fig:S1phaseMap} shows the phase obtained from Eq. (\ref{eq:wavefront}). Figure \ref{fig:S1fringes} shows the fringe patterns $I^{c}_{\mathbf{x}}$ and $I^{s}_{\mathbf{x}}$ (Eq. (\ref{eq:phaseShift})) used in the estimation, with resolution of 640 $\times$ 480 pixels. A first experiment was the phase estimation using the fringe patterns shown in Figure \ref{fig:S1fringes}, where $\lambda = 1$ and the value for $\phi^{0}_{\mathbf{x}}$ was randomly generated. The estimated phase map is shown in Figure \ref{fig:S1estimacion}. The normalized error was $Q = 0.0014$ and the time employed to obtain the solution was 106 seconds using 2342 iterations of the Nesterov’s method (Eq. (\ref{eq:Nesterov})). Figure \ref{fig:S1errorEst} shows the error obtained in this estimation.

\begin{figure}[ht!]
  \begin{center}
    \includegraphics[width=\textwidth]{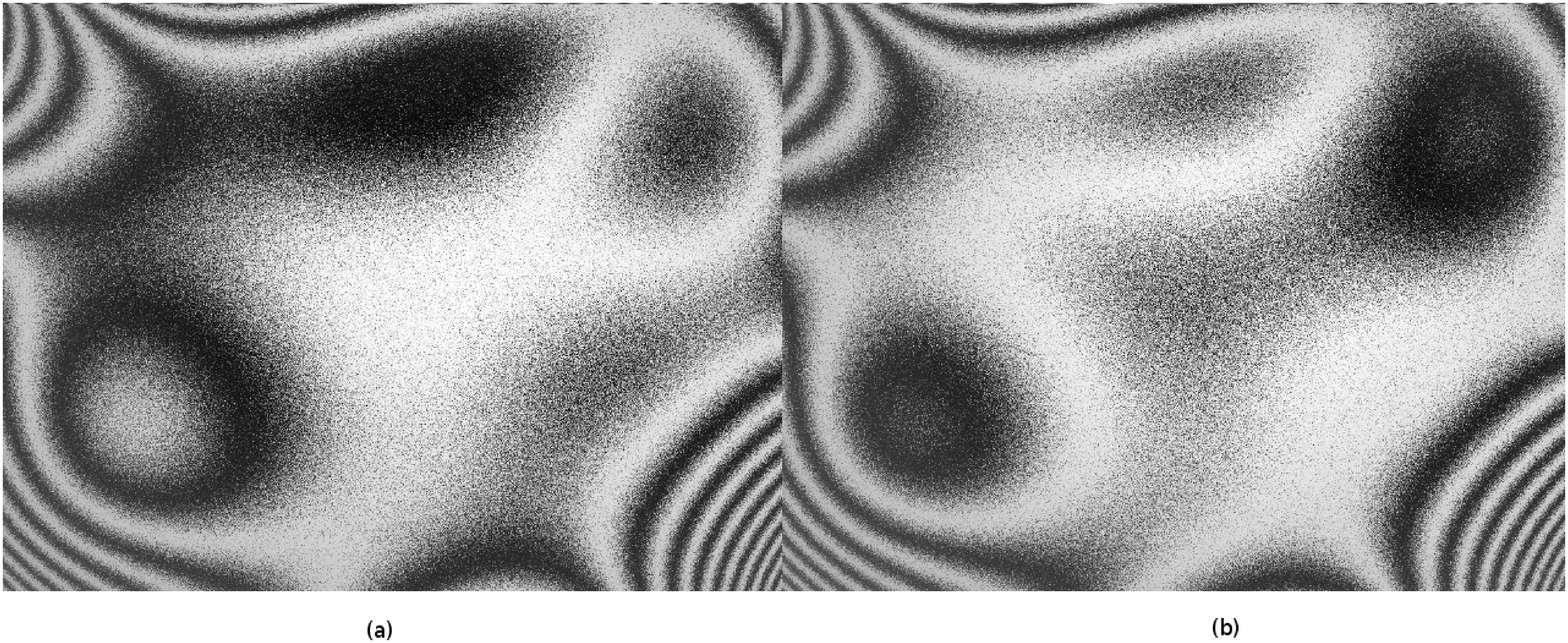}
    \caption{Noisy fringe patterns generated with the phase map given in Eq. (\ref{eq:wavefront}): (a) $I^{c}_{\mathbf{x}}$, (b) $I^{s}_{\mathbf{x}}$.}
    \label{fig:S2fringes}
  \end{center}
\end{figure}

\begin{figure}[ht!]
  \begin{center}
    \includegraphics[width=\textwidth]{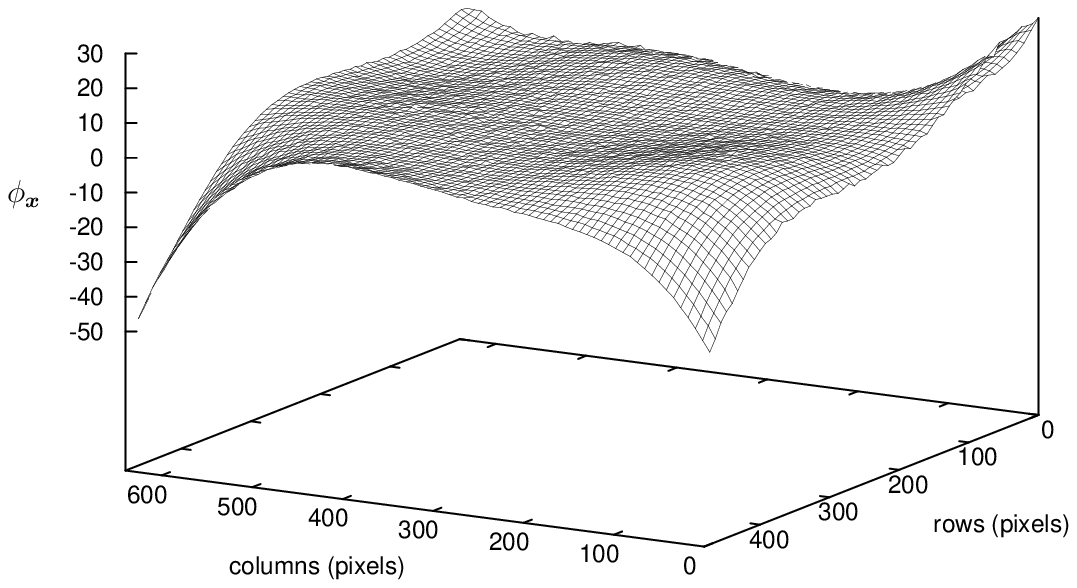}
    \caption{Estimated phase map using fringe patterns shown in Fig. \ref{fig:S2fringes}.}
    \label{fig:S2estimacion}
  \end{center}
\end{figure}

\begin{figure}[ht!]
  \begin{center}
    \includegraphics[width=\textwidth]{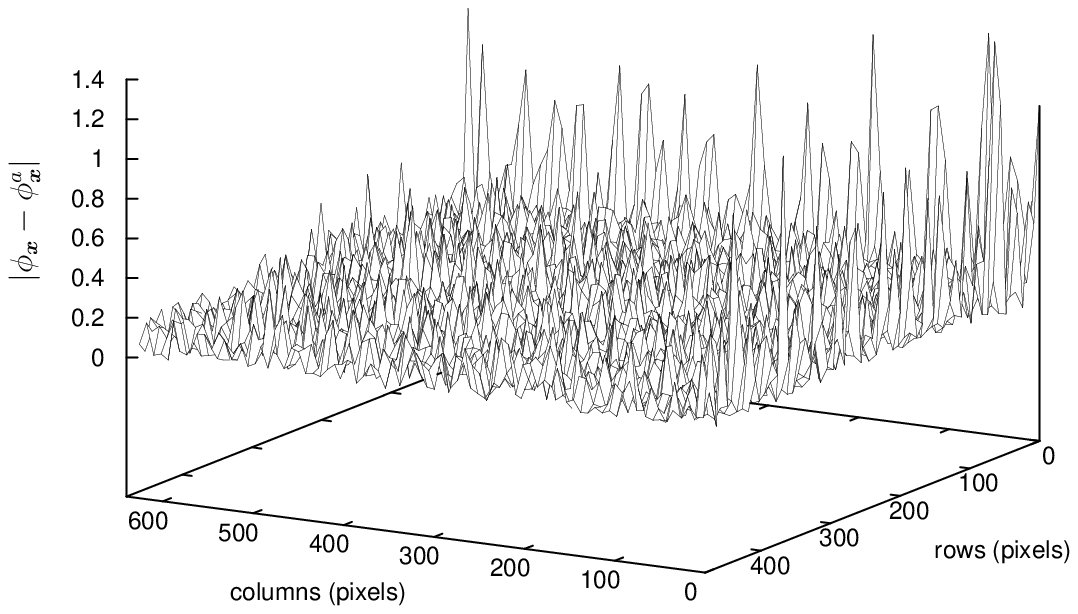}
    \caption{Absolute difference between the estimated phase map using fringe patterns shown in Fig. \ref{fig:S2fringes} and the synthetic phase map given in Eq. (\ref{eq:wavefront}).}
    \label{fig:S2errorEst}
  \end{center}
\end{figure}

A second estimation was made using fringe patterns with SNR = 12.5 db~\cite{Gonzalez2008}, shown in Figure \ref{fig:S2fringes}. For this estimation, we use $\lambda = 1.0$ and the initial value was randomly generated. The estimated phase map is shown in Figure \ref{fig:S2estimacion}. The normalized error was $Q = 0.016$ and the time employed to obtain the solution was 123 seconds using 2819 iterations of the Nesterov’s method. Figure \ref{fig:S2errorEst} shows the error obtained in this estimation. Table \ref{Table:surfaceEstimation} presents a summary of the estimation performance of our proposed functional using the fringes patterns generated by Eq. (\ref{eq:wavefront}) with different SNR.

\begin{figure}[ht!]
  \begin{center}
    \includegraphics[width=\textwidth]{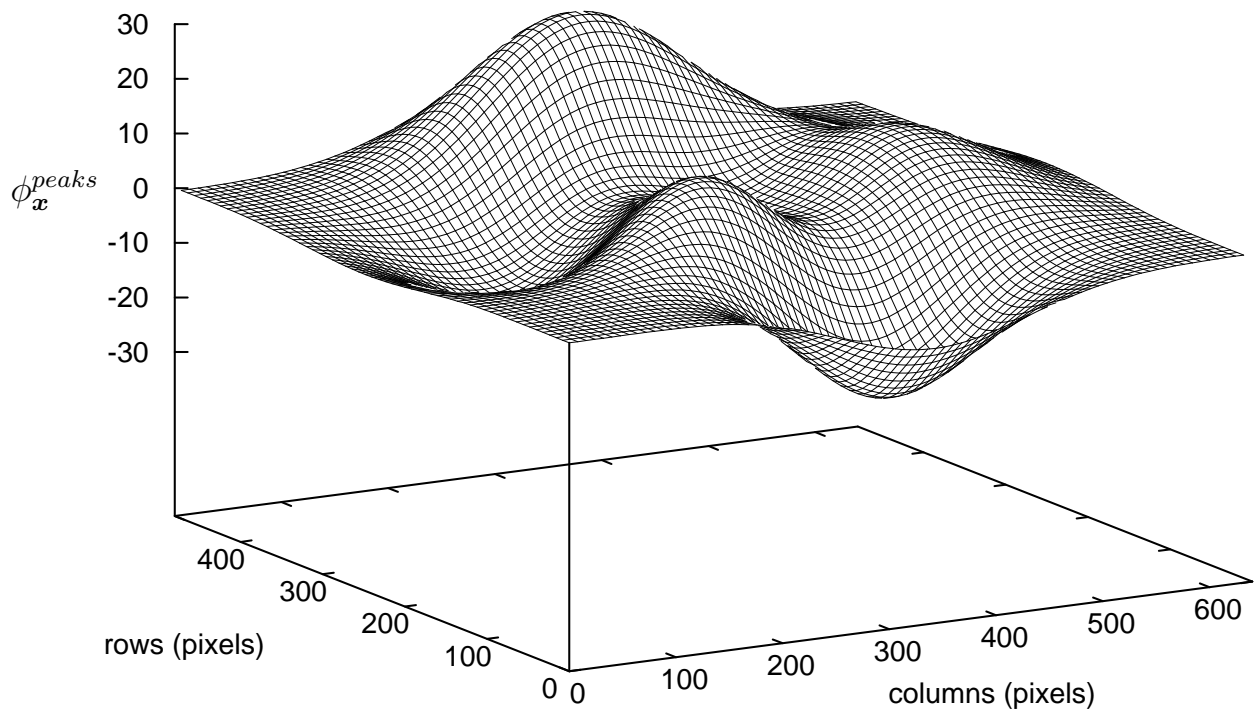}
    \caption{Synthetic phase map given by \texttt{peaks} function.}
    \label{fig:S3phaseMap}
  \end{center}
\end{figure}

\begin{figure}[ht!]
  \begin{center}
    \includegraphics[width=\textwidth]{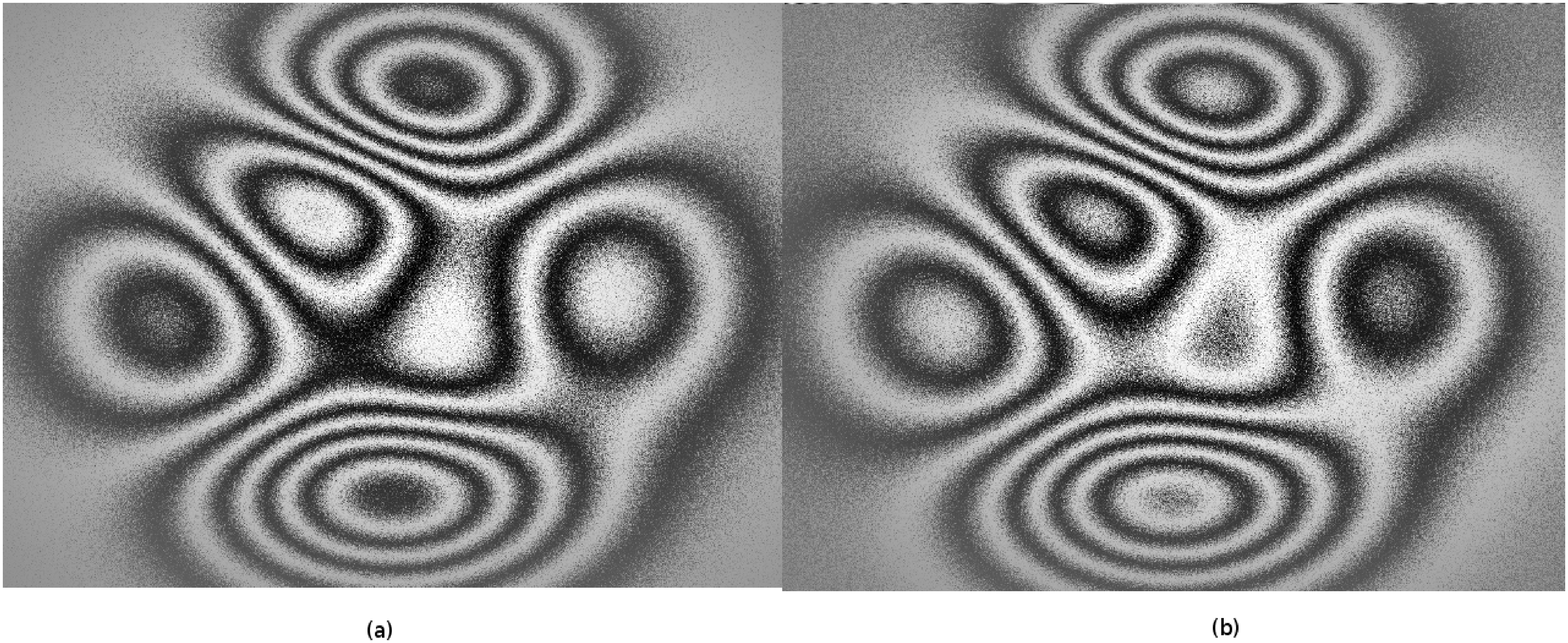}
    \caption{Noisy fringe patterns generated with the phase map given in Fig. \ref{fig:S3phaseMap}: (a) $I^{c}_{\mathbf{x}}$, (b) $I^{s}_{\mathbf{x}}$.}
    \label{fig:S3fringes}
  \end{center}
\end{figure}

\begin{figure}[ht!]
  \begin{center}
    \includegraphics[width=\textwidth]{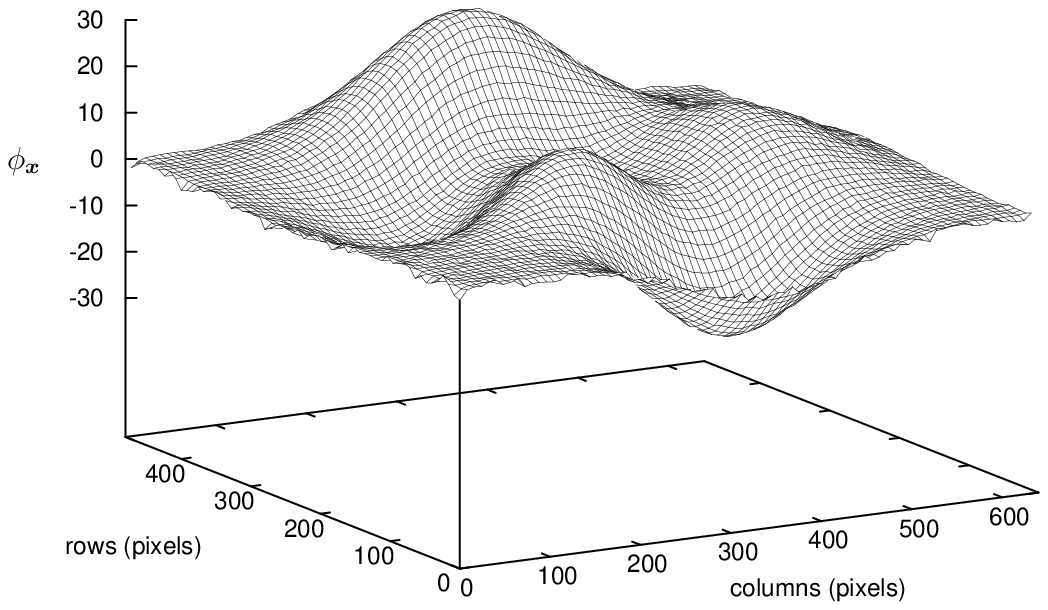}
    \caption{Estimated phase map using fringe patterns shown in Fig. \ref{fig:S3fringes}.}
    \label{fig:S3estimacion}
  \end{center}
\end{figure}

\begin{figure}[ht!]
  \begin{center}
    \includegraphics[width=\textwidth]{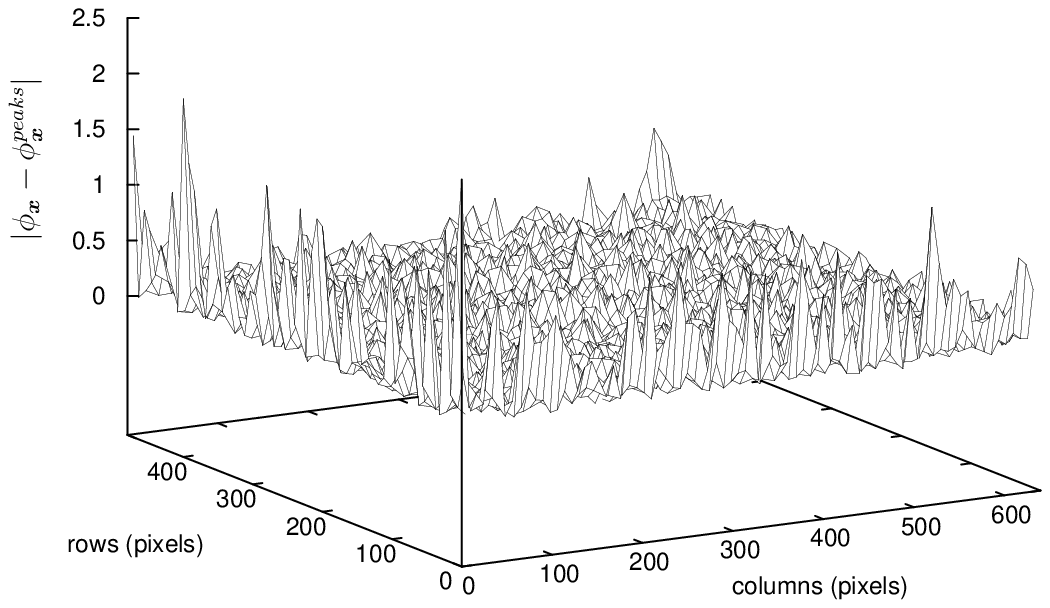}
    \caption{Absolute difference between the estimated phase map using fringe patterns shown in Fig. \ref{fig:S3fringes} and the synthetic phase map given by \texttt{peaks} function.}
    \label{fig:S3errorEst}
  \end{center}
\end{figure}

A second experiment was the estimation of a synthetic phase map defined by the MATLAB \texttt{peaks} function\cite{Matlab}, evaluated in a square domain\\ 
$\Omega = \left\lbrace (x,y)\,|\, -2.3\leq x,y\leq 2.3\right\rbrace.$ Figure \ref{fig:S3phaseMap} shows the wrapped phase generated by the \texttt{peaks} function, and Figure \ref{fig:S3fringes} shows the fringe patterns $I^{c}_{\mathbf{x}}$ and $I^{s}_{\mathbf{x}}$ (Eq. (\ref{eq:phaseShift})) used in the estimation, with resolution of 640 $\times$ 480 pixels with SNR = 12.7 db. The resultant estimated phase map is shown in Figure \ref{fig:S3estimacion} where the normalized error was $Q = 0.014$ and the time employed to obtain the solution was 342 seconds using 7552 iterations of the Nesterov’s method. Figure \ref{fig:S3errorEst} shows the error obtained in this estimation. Table \ref{Table:peaksEstimation} presents a summary of the estimation performance of our proposed functional using the fringes patterns generated by the MATLAB \texttt{peaks} function with different SNR.

\subsection{Phase estimation using experimental fringe patterns}
\begin{figure}[ht!]
  \begin{center}
    \includegraphics[width=\textwidth]{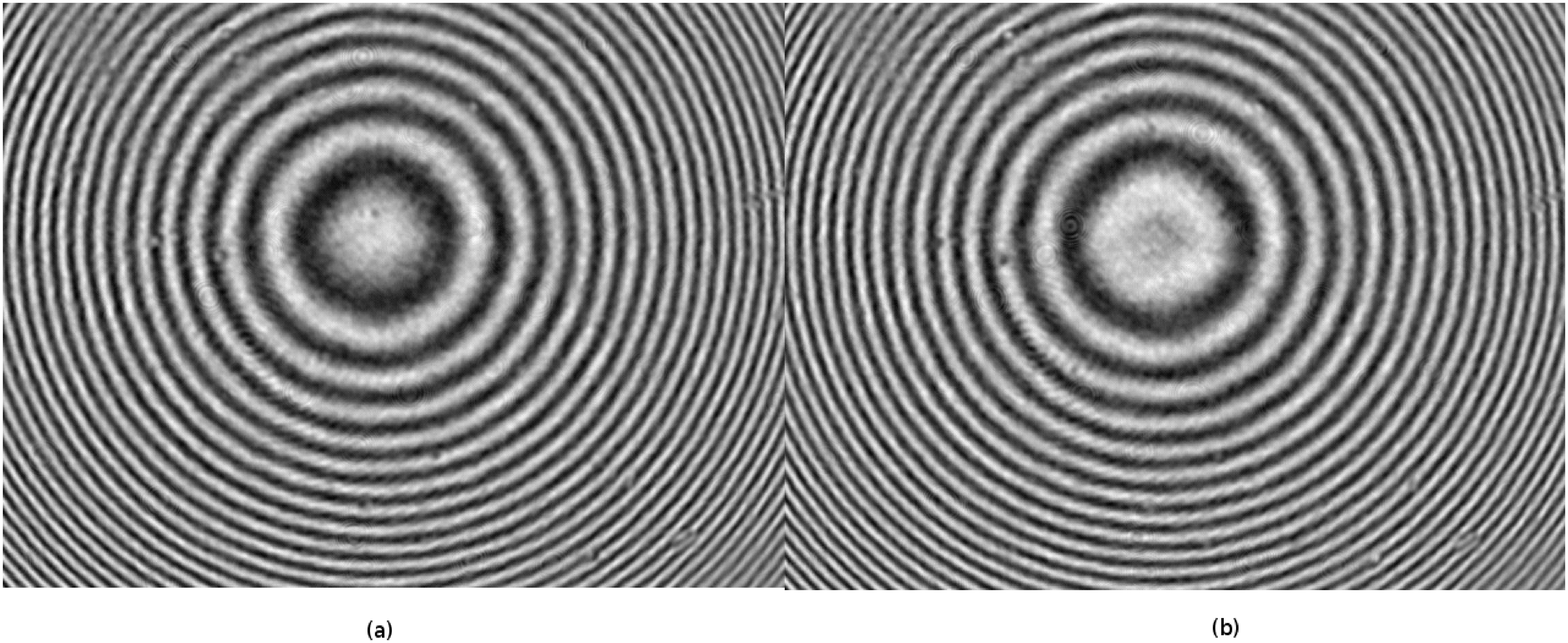}
    \caption{Experimental fringe patterns: (a) $I^{c}_{\mathbf{x}}$, (b) $I^{s}_{\mathbf{x}}$.}
    \label{fig:R1fringes}
  \end{center}
\end{figure}

\begin{figure}[ht!]
  \begin{center}
    \includegraphics[width=\textwidth]{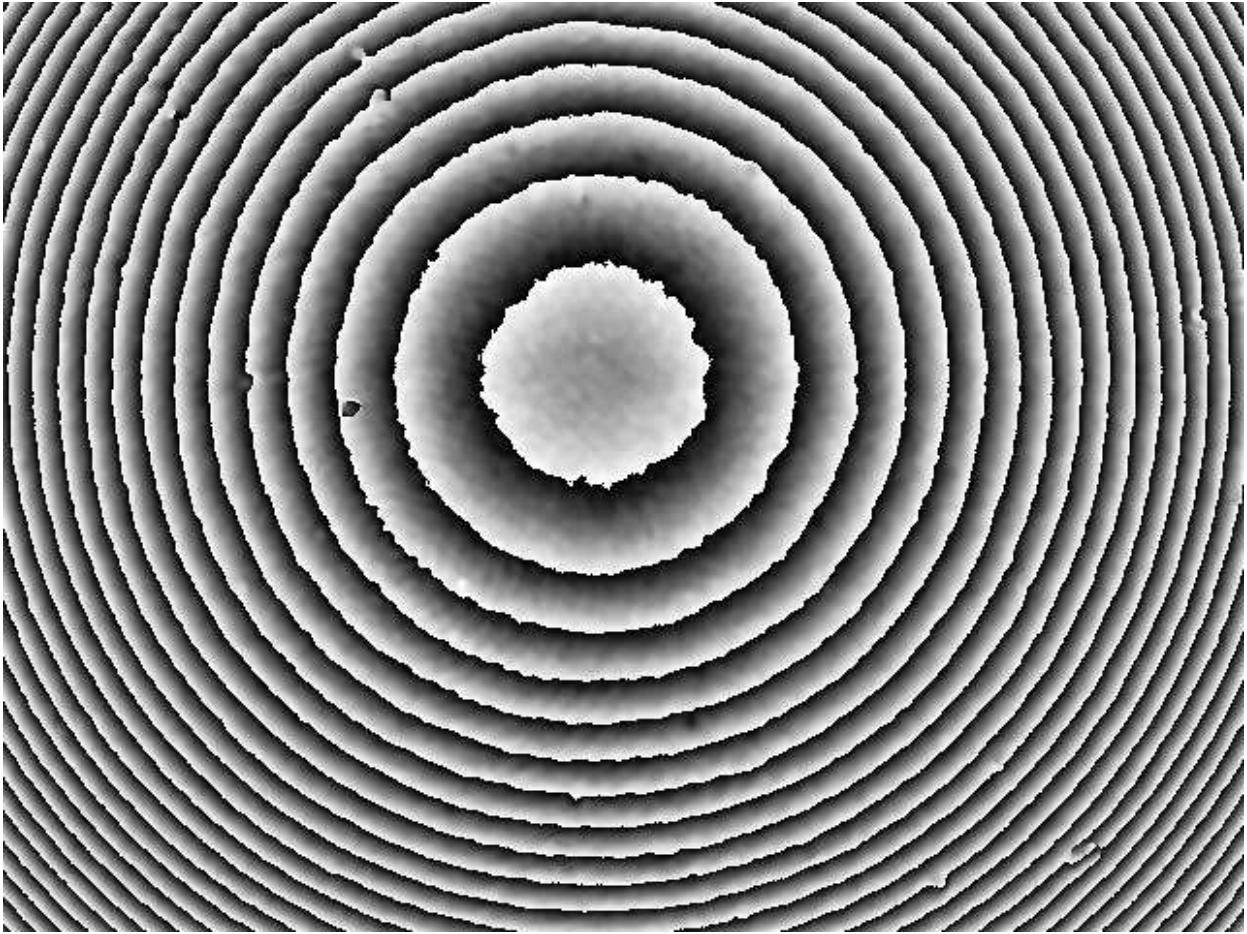}
    \caption{Wrapped phase map generated by the fringe patterns shown in Fig. \ref{fig:R1fringes}}
    \label{fig:R1phaseMap}
  \end{center}
\end{figure}

In this experiment we show the performance of the proposed method on the processing of experimental information with noise. This experiment consists of the phase estimation of a sequence of 5 fringe patterns obtained from a holographic interferometric experiment~\cite{Kreis1996}. Figure \ref{fig:R1fringes} shows the fringe patterns $I^{c}_{\mathbf{x}}$ and $I^{s}_{\mathbf{x}}$ (Eq. (\ref{eq:phaseShift})) obtained from the phase-shifting method, with resolution of 640 $\times$ 480 pixels. The wrapped phase map obtained from these fringe patterns can be observed in Figure \ref{fig:R1phaseMap}. 

\begin{figure}[ht!]
  \begin{center}
    \includegraphics[width=\textwidth]{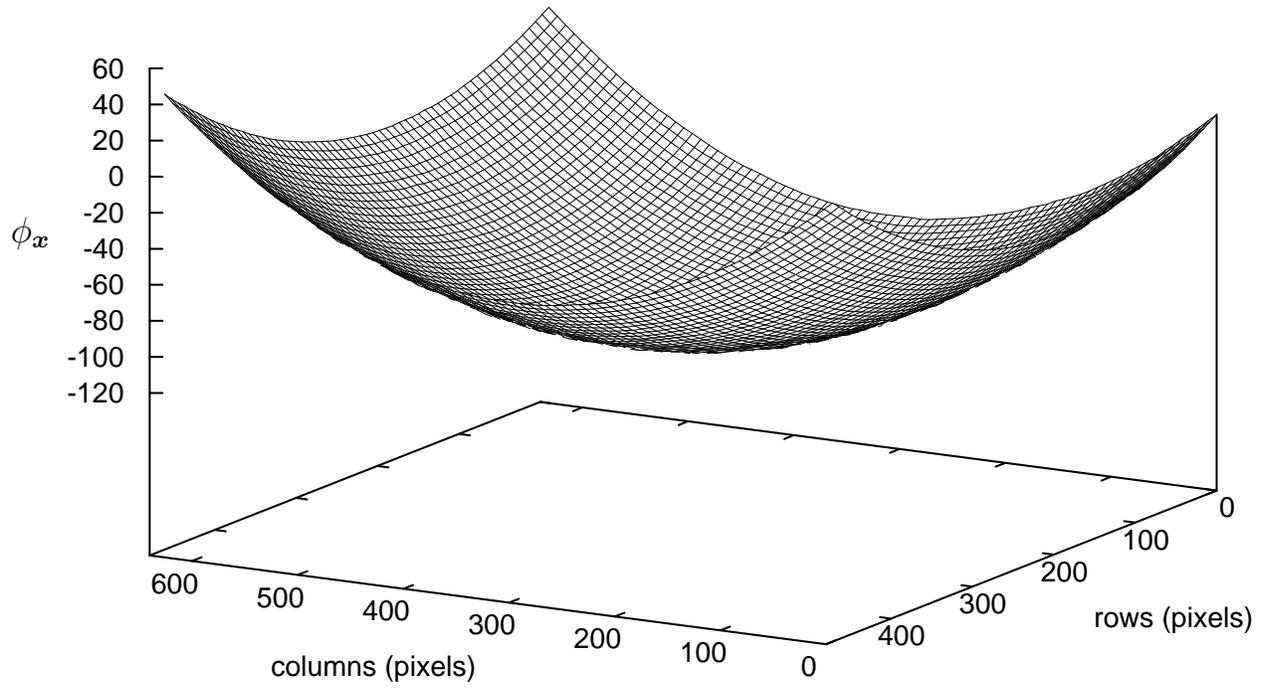}
    \caption{Estimated phase map using Eq. (\ref{eq:energiaMinimiza}).}
    \label{fig:R1estimacion}
  \end{center}
\end{figure}

Due to the noisy phase-term, the strong variations in the modulation or the presence of phase-shift miscalibration~\cite{Surrel1996}, the iterative process will be slow or even trapped on a local minimum. To improve the iterative process, we propose to use as initial value the phase term obtained from the method reported in reference \citenum{Paez1999}. Figure \ref{fig:R1estimacion} shows the estimated phase using $\lambda = 1.5$; the normalized error was $Q = 0.012$ and the time employed to obtain the solution was 87 seconds, including the time employed to estimate the initial value, using 1590 iterations of the Nesterov’s method. 

\begin{figure}[ht!]
  \begin{center}
    \includegraphics[width=\textwidth]{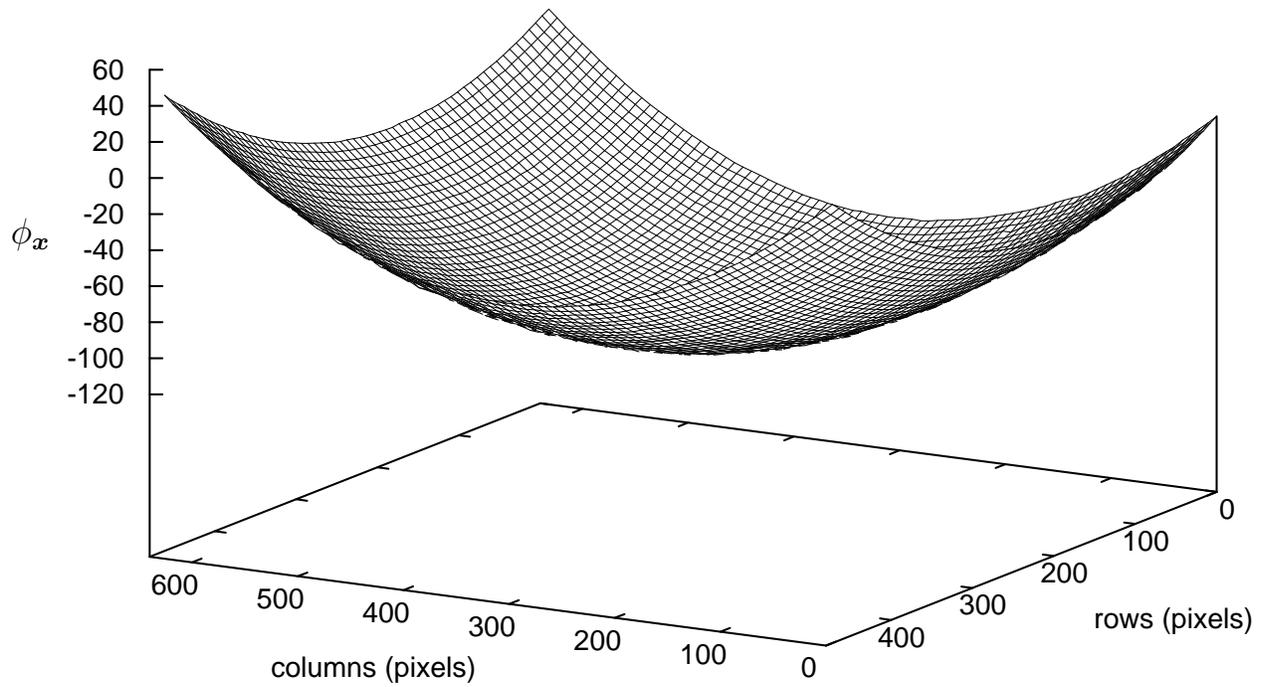}
    \caption{Estimated phase map using Eq. (5.31) of reference \citenum{Ghiglia1998}.}
    \label{fig:R2estimacion}
  \end{center}
\end{figure}

\begin{figure}[ht!]
  \begin{center}
    \includegraphics[width=\textwidth]{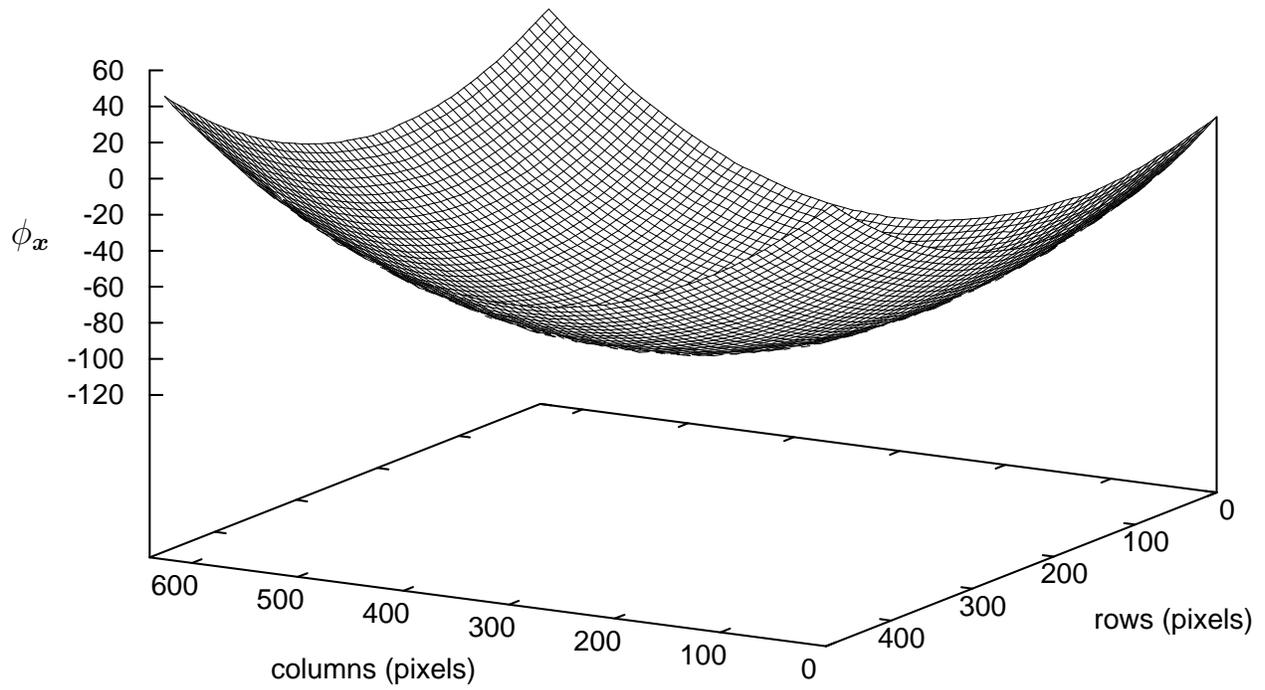}
    \caption{Estimated phase map using reference \citenum{Marroquin1995a}.}
    \label{fig:R3estimacion}
  \end{center}
\end{figure}

To compare the performance of our proposal, we unwrap the phase map shown in Figure \ref{fig:R1phaseMap} with two unwrapping methods: a) the discrete version of Poisson equation (Eq. (5.31) of reference \citenum{Ghiglia1998}), and b) the method described in reference \citenum{Marroquin1995a}. For both cases, we used the Nesterov’s method as optimization technique. Figure \ref{fig:R2estimacion} shows the estimated phase using reference \citenum{Ghiglia1998}, where the time employed to obtain the solution was 40 seconds, including the time employed to estimate the initial value, using 5205 iterations of the Nesterov’s method and the normalized error was $Q = 0.015$. Figure \ref{fig:R3estimacion} shows the estimated phase using reference \citenum{Marroquin1995a} with $\lambda = 1.2$, where the time employed to obtain the solution was 245 seconds, including the time employed to estimate the initial value, using 15000 iterations of the Nesterov’s method and the normalized error was $Q = 0.241$.

\section{Discussion of results and conclusions}
As can be observed from the above experiments, the proposed method successfully estimates the unwrapped phase map from the information of the phase-shifting method methods; that is, the fringe patterns $I^{c}_{\mathbf{x}}$ and $I^{s}_{\mathbf{x}}$, without the use of the wrapped phase map. This method converges to an accurate solution given an arbitrary initial point, even for noisy fringe patterns. Due to the smoothing term included in the functional, it is possible to obtain a filtered phase map with the preservation of the dynamic range of the fringe patterns. 

The numerical solution of Eq. (\ref{eq:energiaMinimiza}) results on a very simple algorithm which estimate a filtered phase map in a short time, despite the optimization algorithm with poor convergence-rate used in experiments. In comparison with the methods described in reference \citenum{Ghiglia1998} (Eq. (5.31) and reference \citenum{Marroquin1995a}, the proposed method shows a good performance on the estimation and the computational load employed to compute the phase map. To improve the convergence-rate, the proposed method can be easily implemented with a better computationally efficient techniques, including parallel approaches. This will be one aim of our future research.

\newpage
\bibliographystyle{tfs}

\newpage
\begin{table}[ht]
  \centering
  \caption{Estimation performance of Eq. (\ref{eq:energiaMinimiza}) using fringes patterns generated with Eq. (\ref{eq:wavefront}).}\label{Table:surfaceEstimation}
  \begin{tabular}[t]{ccc}
  \hline
  SNR (db)& iterations&   Normalized error (Q)\\
  \hline
  $\inf$ & 2342 & 0.00142\\
  39.97 &  2428 & 0.00335 \\
  27.98 &  2483 & 0.00980\\
  21.08 &  2501 & 0.00867\\
  \hline
  \end{tabular}
\end{table}

\begin{table}[ht]
  \centering
  \caption{Estimation performance of Eq. (\ref{eq:energiaMinimiza}) using fringes patterns generated with \texttt{peaks} function.}\label{Table:peaksEstimation}
  \begin{tabular}[t]{ccc}
  \hline
  SNR (db)& iterations&   Normalized error (Q)\\
  \hline
  $\inf$ & 5555 & 0.00072\\
  40.26  & 5876 & 0.00366\\
  28.22  & 6350 & 0.00603\\
  21.22  & 6782 & 0.00823\\
  14.44  & 7349 & 0.01076\\
  12.72  & 7552 & 0.01427\\
  \hline
  \end{tabular}
\end{table}

\newpage
\section*{Figures}

\end{document}